\documentstyle[11pt,epsf,epsfig]{article}
\begin{document}
\unitlength=1cm

\begin{flushright}
LAPTH 692-98 \\
UM-TH 98-12  \\
CERN-TH/98-244  \\
August 1998
\end{flushright}
\vspace{2.cm}
\begin{center}

\large\bf

{\LARGE\bf New developments in the $1/N$ expansion and nonperturbative Higgs physics}\\[2cm]
\rm
{ Thomas Binoth$^a$ and Adrian Ghinculov$^{b,c,}$\footnote{Work supported by the 
                                   US Department of Energy (DOE)}}\\[.5cm]

{\em $^a$Laboratoire d'Annecy-Le-Vieux de Physique 
         Th\'eorique\footnote{URA 1436 associ\'ee \`a l'Universit\'e de Savoie} LAPP,}\\
      {\em Chemin de Bellevue, B.P. 110, F-74941, 
           Annecy-le-Vieux, France}\\[.2cm]

{\em $^b$Randall Laboratory of Physics, University of Michigan,}\\
      {\em Ann Arbor, Michigan 48109-1120, USA}\\[.2cm]

{\em $^c$CERN, 1211 Geneva 23, Switzerland}\\[2.cm]

\end{center}
\normalsize

\begin{abstract}
We show in this paper that the $1/N$ expansion is a reliable tool
to calculate the properties of a heavy Higgs boson. 
The 1/N expansion sums up all orders in perturbation theory,
and therefore avoids  the renormalization scheme dependence
of the conventional perturbative approach.
It is explained how effects due to the Landau pole of the Higgs sector
are isolated and subtracted, and how to perform 
actual calculations, by computing
the Higgs line shape for the processes $f\bar f\rightarrow H \rightarrow ZZ,f'\bar f'$
at next-to-leading order in the $1/N$ expansion. 
The results are compared to the perturbative results to
show the agreement between the perturbative and 
the nonperturbative approach for Higgs masses up to 1 TeV.
We conclude that the theoretical predictions for Higgs observables
are well under control for the entire kinematical region of the LHC.      
\end{abstract}


\section{Introduction}

A main goal of future collider physics is to shed light into the
electroweak symmetry breaking mechanism. Within the standard model
this mechanism is generated by the Higgs sector, leading to gauge 
invariant mass terms where needed. 
At the Lagrangean level, this is a gauged linear $O(4)$-symmetric
sigma model, spontaneously broken to $O(3)$ by a nonvanishing vacuum
expectation value.
The mass of the Higgs boson
itself is directly related to the corresponding quartic coupling,
which implies the nondecoupling property of the Higgs boson.
For large Higgs mass one expects large quantum corrections which 
blow up within the standard perturbative approach.

Within the perturbative approach, 
a lot of work has been done in the recent years. 
Many quantum corrections for relevant observables which
contain information on the Higgs boson
are available \cite{kniehl:higgs1loop}. 
The radiative corrections 
of enhanced electroweak strength to the decay of the Higgs boson
were first derived at one-loop  
in \cite{marciano}, and at two-loop in
\cite{ghinculov:2loop,maher:2loop,jikia:2loop}. 
For low values of the Higgs mass the theoretical uncertainty
is thus small. For heavier masses the leading order and
next-to-leading order results, 
start to deviate from the NNLO result, inducing theoretical ambiguities. 
One expects large dependences on
the renormalization scheme essentially coming from the truncation
of the perturbative series which makes theoretical predictions unreliable. 
Whereas the perturbative region will be fully covered
by LEP, Tevatron and the LHC, it is of special 
importance to have a good theoretical understanding of the heavy Higgs
regime which is also in reach of the LHC. 
Although the existing experimental data points presently in the 
direction of a Higgs mass in the perturbative region,
one should not forget that these bounds are logarithmically soft,
and that the experimental data on
$\sin^2 \theta_{eff}$ still differ considerably in the high precision
measurements at LEP1 and SLC \cite{degrassi:higgsbound}.
Moreover, certain two-loop quantities are presently known theoretically
only as mass expansions, and given the high statistics of the experiments,
this induces rather substantial shifts in the Higgs mass prediction.

Apart from the phenomenological motivation, it is a major challenge to
develop and improve calculational methods beyond the standard 
perturbative approach. A well-known nonperturbative 
approach are Monte Carlo studies on a
lattice. The results of lattice calculations indicate
that for Higgs masses beyond $\sim$ 700 GeV 
one expects large cutoff effects \cite{lattice:higgsbound}. 
They stem from the fact
that the Higgs sector is not asymptotically free. 
We advocate another nonperturbative approach, the $1/N$ expansion.
There, the quantum correction are ordered by the number of field components 
\cite{coleman,schnitzer}. The method was applied to the Higgs sector 
to leading order a long time ago \cite{einhorn} but the large deviation 
from standard perturbative results shaded doubts on its validity.

Recently we performed a next-to-leading order 
calculation of the Higgs propagator, showing at low coupling a remarkable 
agreement between the NNLO perturbative result and our result \cite{1on:nlo}.
This proved the $1/N$ expansion to be a valuable tool for 
computing observables beyond the perturbative approach, and for 
avoiding the truncation of the perturbative series and thus 
eliminating any renormalization scale dependence.

In this paper we extend the results to the case of three-point
functions, which results in a quantitative, nonperturbative
understanding of the physical processes 
$f\bar f \rightarrow H \rightarrow f'\bar f',Z_L Z_L$.
It was tried to make the computation as transparent as possible
to convince the reader that the method is conceptually well settled and
practicable, though some amount
of numerical work is unavoidable, which on the other hand is common to the
calculation of higher-order finite momentum 
Green functions in  massive theories.

After some preliminaries, where we remind the LO results and the 
issue of the tachyon of the $1/N$ approach, 
we show how one obtains a well defined,
tachyon free representation of Green functions. An ultraviolet
subtraction scheme is defined in terms of 
an order $1/N$ counterterm Lagrangean.
Then, the finite and tachyon free Green functions are expressed 
in form of multiloop
graphs made out of dressed propagators.
These can be reduced to two-dimensional integrals 
over dressed propagators and form factors.
Then we describe how to perform the remaining integrations
numerically, to obtain the final results for the 
two- and three-point functions.
We show and discuss our nonperturbative result 
for the Higgs line shapes in two physical scattering processes,
and compare to the perturbative result.
After this discussion, we summarize the conclusions of the paper.


\section{Preliminaries}

In this section we review the leading order result for
fixing our notations, and discuss the structure of the 
counterterms which absorb the ultraviolet
divergencies at leading and next-to-leading order.


Because we are interested in the effects of a heavy Higgs
boson, we neglect all the gauge couplings and also the higher order 
corrections in the Yukawa couplings.
Thus, the starting point for the $1/N$ expansion in a scalar theory
is given by an $O(N)$-symmetric $\phi^4$ theory:

\begin{eqnarray}\label{lag1}
{\cal L}= \frac{1}{2}\, \partial_{\nu} \vec\phi_0 \partial^{\nu} \vec\phi_0 - 
          \frac{1}{2} \mu^2_0 \, \vec\phi_0^2 - 
          \frac{\lambda_0}{4! N} (\vec\phi_0^2)^2
\end{eqnarray}
Here $\vec\phi_0$ is a $N$-component vector and $\mu_0$, $\lambda_0$ are  
bare coupling constants indicated by the subscript zero.
Because we want to focus on the spontaneously broken case, we rewrite it as   

\begin{eqnarray}\label{lag2}
{\cal L}= \frac{1}{2}\, \partial_{\nu} \vec\phi_0  \partial^{\nu} \vec\phi_0 
         - \frac{\lambda_0}{4! N} (\vec\phi_0^2 - N v_0^2)^2
\end{eqnarray}
$N v_0^2=-6 N \mu_0^2/\lambda_0$ is the bare vacuum expectation value of 
$\vec\phi^2$.
The symmetry of the ground state is broken down to $O(N-1)$, which 
amounts to the presence of $N-1$ massless Goldstone bosons (GBs), 
$\pi_{i=1,\dots,N-1}$,
and a Higgs particle $h$, of mass $m_h^2=\lambda_0v_0^2/3$. 
In the case $N=4$,
the GBs are related to the longitudinal degrees of freedom of the 
vector bosons $W_L$, $Z_L$
by the equivalence theorem.

To overcome the combinatorial difficulties of multi-loop 
calculations, we use a well-known
auxiliary field formalism \cite{coleman}.
We add to the Lagrangean the following 
expression, which becomes identically zero
when one uses the equations of motion:

\begin{eqnarray}\label{lag_aux}
{\cal L_{\chi}} = \frac{3N}{2\lambda_0} Z_\kappa \Bigl[ \chi_0 - 
\frac{\lambda_0}{6N} ( \vec\phi_0^2 - N v^2_0 ) \Bigr]^2 . 
\end{eqnarray}
$Z_{\kappa}$ is an arbitrary Lagrange multiplier. One additionally has 
the freedom to replace 
$\chi_0$ by $\alpha \chi_0 + \beta$, where $\alpha, \beta$
are some constants, without changing the equations of motion.
These constants correspond to a $\chi$ vacuum expectation value
and a wave function renormalization. 
We introduce renormalized quantities in the
following way:

\begin{eqnarray}\label{renor}
\phi_0=\sqrt{Z_{\phi}}\,\phi=\sqrt{Z_{\phi}}(\vec\pi,\sqrt{\frac{Z_h}{Z_\phi}}\sigma), 
\quad v_0=\sqrt{Z_{\phi}}Z_v v, \quad \lambda_0=\frac{Z_\lambda}{Z_{\phi}^2} \lambda, 
\quad \chi_0=\frac{Z_{\chi}}{Z_{\phi}}(\chi+V_\chi) 
\end{eqnarray}

From the discussion above it is clear that only $Z_\phi$, $Z_h$, $Z_\lambda$
and $Z_v$ are renormalization constants in the usual sense.
The other can be chosen arbitrarily as long as the choice is
consistent.
Their introduction will turn out to be useful for the 
renormalization of functions which contain auxiliary particles
as external legs. 
The Lagrangean is now of the form:

\begin{eqnarray}  \label{lagrenor}
{\cal L} &=& \frac{c_0}{2} \partial_{\nu} \vec\phi \partial^{\nu} \vec\phi 
        + c_1 Nv^2\chi + \frac{c_2}{2}\frac{Nv^2}{m_h^2} \chi^2 \nonumber \\
     & &  - \frac{c_3}{2} \chi ( \vec\phi^2 - N v^2 ) - \frac{c_4}{2} (\vec\phi^2 - N v^2) - 
            \frac{c_5}{8}\frac{m_h^2}{Nv^2} (\vec\phi^2 - N v^2)^2 \nonumber\\
      \mbox{with}\qquad && \nonumber \\
\vec\phi&=&( \vec \pi, c_6\, \sigma+\sqrt{N} v) \quad .       
\end{eqnarray}

Note that a mixing between $\chi$ and $\sigma$ takes place.
We see that the expression 
contains --- up to a trivial surface term $\Box \chi$ ---
all possible field combinations of dimension four which are
built out of an $O(N)$ vector and a dimension two singlet 
field allowed by
Lorentz invariance and the $O(N)$ symmetry. The vacuum expectation value
of the Higgs reduces the symmetry as usual to $O(N-1)$.
The constants $c_{i=0,\dots,6}$ are expressible 
in terms of the above renormalization
constants. 

\begin{eqnarray}\label{cis}
\begin{array}{lll}
c_0 = Z_{\phi} & c_1 = Z_\chi Z_{\kappa} ( \frac{Z_\chi V_\chi }{Z_\lambda m_h^2} 
         + \frac{Z_v^2-1}{2} ) & c_2 = \frac{Z_{\kappa} Z_\chi^2}{Z_\lambda}  \\
c_3 = Z_\chi Z_{\kappa} & c_4 =  Z_\chi Z_{\kappa} V_\chi & 
c_5 =  (1-Z_{\kappa})  Z_\lambda \\
c_6 = \sqrt{Z_h/Z_{\phi}} & &
\end{array}
\end{eqnarray}

These constants define a redundant counterterm structure of the theory, 
and are fixed by  imposing conditions on
one-particle irreducible Green functions. 
Whereas some are necessary to absorb the divergencies 
occurring in physical processes, other can be adjusted
to make subtractions on Green functions containing auxiliary fields as
external legs.  
Clearly all these  subtractions have to be consistent with each other.
In order to determine the renormalization constants, the quantum corrections 
to the tree level Green functions have to be calculated up to a given order,
and have to be related to the counterterms  by a renormalization scheme. 

In perturbation theory the occurring renormalization scheme dependence 
is proportional to some power of the coupling constant, and namely
one order higher than the loop order in which the process is calculated. 
Thus it is enhanced in the case of strong coupling. 
It is a crucial point that this is not the case in the $1/N$ expansion. 
Once the relations between 
a set of observables and the Green functions are fixed, all further 
predictions are exact, up to unevaluated powers of $1/N$.
A change of the ultraviolet subtraction point $\mu^2$ (see below) 
is compensated
by a change in $\lambda$. Thus it does not effect any observable prediction, 
being merely a reparameterization of the theory.

Before describing the computation
of the Higgs propagator in next-to-leading order in $1/N$, we would like to 
repeat the leading order results to set the stage for the next step.  
To order $(1/N)^0$ the  
quantum corrections are simply the sum of all graphs containing 
arbitrary numbers of external $\chi$ legs attached to a GB loop.
The sum can be carried out and a simple closed expression for
the resulting effective potential can be found \cite{coleman}.
The only divergent graphs are 
a $\chi$ tadpole and a momentum dependent $\chi$ self-energy graph
shown in figure 1.
\begin{figure}
\begin{center} 
\setlength{\unitlength}{1.0cm}
\begin{picture}(13,2.5) \put(+1.5,+1.25){\makebox(0,0){$T_\chi^0=\frac{1}{N-1}$}}
                       \put(+2.5,+0.25){\epsfxsize3cm  \epsfysize2cm  \epsffile{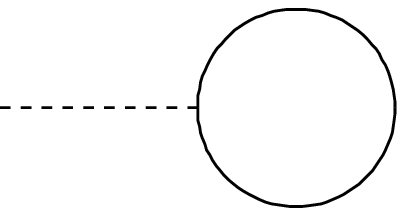} }
                       \put(+6.5,+1.25){\makebox(0,0){,$\quad A^0=\frac{1}{N-1}$}}
                       \put(+8.0,+0.25){\epsfxsize4cm  \epsfysize2cm  \epsffile{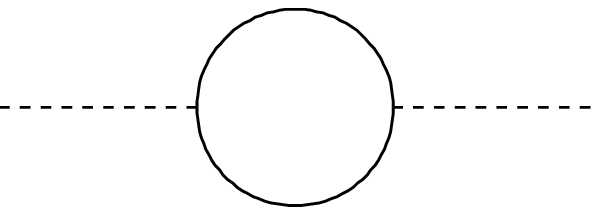} } 
\end{picture}
\end{center} 
\caption{{\em Leading order tadpole and self-energy graphs.}}
\end{figure}
Both divergencies can be absorbed into
$Z_v$ and $Z_{\lambda}$, or in other words, into the
coupling constants of the theory. 
No other renormalization constant is involved.
Whereas the observable vev is 
defined by the ground state of the theory 
--- resulting in an observable vector boson mass if one couples the model to
a gauge theory --- the value of $\lambda$ can be fixed by a 
measurement of the Higgs mass. 
In terms of the introduced renormalization constants, one has:
\begin{eqnarray}\label{loren}
Z_v^2 = 1 - 2\,T_{\chi}^0,\quad 1/Z_{\lambda} = 1 - \lambda A^0(-\mu^2) /3,\quad
Z_{\phi}=Z_h=Z_{\chi}=Z_{\kappa}=1, \quad  V_{\chi}=0 \quad .
\end{eqnarray}
The integrals which correspond to the leading order corrections are trivial.
Absorbing combinatorial factors in the expressions, one finds in dimensional
regularization:
\begin{eqnarray}
A^0(s)     &=& -\frac{1}{32\pi^2} \log(-s/\Lambda^2) + 
                   \frac{\mbox{const.}}{\epsilon} + \mbox{const.'} \nonumber \\
T_{\chi}^0 &=&  0 
\end{eqnarray}
Defining an energy dependent Higgs mass by $m^2(s) = v^2\lambda(s)/3$,
where $1/\lambda(s)=1/\lambda + (\alpha_0(s)-\alpha_0(\mu^2))/3$,
one gets an expression for the matrix of two-point functions.
Its inverse defines the dressed propagators to leading order. 
$\mu$ is the renormalization scale.
The resulting propagators  are found to be:
\begin{eqnarray}\label{propa}
 D_{hh} &=& \frac{i}{ s - m^2(s) } \nonumber \\
 D_{h\chi} &=& \frac{i}{\sqrt{N}v} \frac{m^2(s)}{ s - m^2(s) }   \nonumber\\
 D_{\chi\chi} &=& \frac{i}{Nv^2} \frac{s\,m^2(s)}{ s - m^2(s) }  \nonumber\\
 D_{\pi_l\pi_k} &=&  \frac{i}{ s }\, \delta_{lk}
\end{eqnarray}  
The GB propagator is not influenced at leading order.
Clearly the same results can be obtained without the auxiliary field. 
The dressed
$\chi$ propagator corresponds to a geometric 
series of loop graphs \cite{einhorn}.
Note that the propagators contain an 
imaginary part for physical, timelike momenta,  
which describes Higgs decay into Goldstone bosons. 
To keep the pole on the correct Riemann 
sheet one has to use the $i\epsilon$ prescription
in the logarithm of the $s$-dependent mass:
 \begin{eqnarray}\label{epspre}
\frac{1}{m^2(s)} = \frac{1}{m^2_h} - \frac{1}{32\pi^2v^2} \log (-\frac{s}{\mu^2}-i\epsilon) 
\end{eqnarray}
For spacelike momenta, a Landau pole occurs at the scale:
\begin{equation}
\Lambda_{L} = \mu \exp\left[ \left(\frac{4\pi v}{m_h}  \right)^2\right]  
\quad .
\end{equation}
Since $m^2(s)$ is negative beyond the Landau scale $\Lambda_L$, all these
propagators have an extra pole for a spacelike 
momentum $s = -\Lambda^2_{T(achyon)}$
which is defined by the equation:
\begin{eqnarray}\label{tachyonscale}
\Lambda_T^2 + m^2(-\Lambda_T^2) = 0\, ,\qquad \Lambda_T^2>0\qquad .
\end{eqnarray}    

There is a tendency to discard non asymptotically free theories
as fundamental theories. The occurrence of the tachyon in the
sigma model is widely believed to be an indication of an ill-defined
quantum field theory. However, to or knowledge there is so far 
no rigorous proof thereof. We will show in the next section that
the tachyon is technically an artifact of perturbation theory and 
can by no means be considered an inconsistency of the model.
Apart from these issues, it is natural to ask for higher order effects.
How to go beyond the leading order in $1/N$ is directly related to the
tachyon problem, because on has to integrate over graphs
built out of tachyon contaminated propagators. 
A method to treat this is given below.
Such a  procedure is touching the aspects of the theory
which cannot be investigated by Feynman diagrams, even
summed up to all orders.
In this sense any tachyonic regularization is modeling either 
effects from a more fundamental theory containing the Higgs sector 
as a low energy limit,
or a certain type of nonperturbative effects within the Higgs sector itself.   
Our nonperturbative treatment of the sigma model is independent of an 
interpretation either as an effective theory or as a fundamental theory,
and we will make no assumption in this direction in this paper.

\section{Calculation}

In this section we explain the necessary steps to calculate
the NLO corrections to the processes 
$f\bar f\rightarrow H \rightarrow Z_L Z_L, W_L W_L, f'\bar f'$
in the $1/N$ expansion. In the case of longitudinal vector boson
decay modes, we calculate the processes by using the equivalence
theorem. Therefore we repace the longitudinal vector bosons
by the corresponding would-be Goldstone bosons $z$ and $w$. 
To this end one has to know the 
two- and three-point functions to that order. 
To get a physically meaningful quantity one has to define a perturbative
renormalization procedure to deal with the usual divergencies.
On the other hand one has to define a scheme to treat the tachyon which
is a perturbative artifact of the leading order calculation. 
We have seen that the infinite series of Feynman diagrams shows up in the propagators
as a momentum dependent mass producing a tachyon pole. 
The dressed propagators in eq. \ref{propa}
are, together with the occurring vertices and the 
counterterms defined in eqn. \ref{lagrenor},
the  building blocks of the diagrams which describe the 
next-to-leading order
effects. To leading order, only GB propagators appear 
as internal lines. 
It is crucial for the simplicity of the calculation that, 
with the help of the auxiliary field $\chi$, the 
next-to-leading order contributions
of the $1/N$ expansion to the Higgs propagator can be depicted 
by a small number of graphs.

In the following, a tachyonic regularization will be defined. 
Then the diagrammatical
part of the calculation will be given together 
with the renormalization procedure.
The subtractions will be explained in detail 
for the relevant multi-loop graphs, and
the finite, tachyon-free quantum corrections to the
two- and three-point functions will be derived. They will be represented as 
two-fold integrals.
Finally the remaining two dimensional integration
is performed numerically, which gives quantitatively the propagator
and vertex corrections.
 
\subsection{Tachyon subtraction}

Before starting the calculation, we introduce
a tachyon-free representation for the occurring propagators
\cite{1on:nlo}.
Assuming that its occurrence is not due the 
failure of the theory under consideration, 
but of the intermediary expansion technique used \cite{1on:nlo}, 
it is reasonable to
circumvent the ill-defined part by an adequate 
subtraction of the tachyonic pole.
If the $\phi^4$ theory is only describing a low energy effective theory, 
some physics at a higher scale will cure the problem.
It is clear that {\em a priori} 
the tachyon regularization is not unique, 
and one has to pay the price in the form of
an additional ambiguity. We modify the propagators 
in eq. \ref{propa} by the following subtraction of the tachyonic pole,
which is the minimal way of eliminating the tachyonic pole:
\begin{eqnarray}\label{tachsub}
\frac{1}{s-m^2(s)} &\rightarrow& \frac{1}{s-m^2(s)} - \frac{1}{s+\Lambda^2_T}\,\kappa \nonumber\\
\frac{m^2(s)}{s-m^2(s)} &\rightarrow& \frac{m^2(s)}{s-m^2(s)} + 
                                \frac{\Lambda^2_T}{s+\Lambda^2_T}\,\kappa
\end{eqnarray}      
$\kappa = 32\pi^2v^2/(32\pi^2v^2+\Lambda_T^2)$ 
is just the residuum of the tachyonic pole.
To get a quantitative feeling for the size of the subtracted terms we
give some values of the appearing constants in table \ref{tachsize}.
It is clear from eq. \ref{epspre} that only a pair of parameters 
$m_h,\mu$ defines a physical situation. We related $\mu$ to the position
of the Higgs pole in the complex plane, $M_{Pole}-i\Gamma_{Pole}/2$, 
by the relation $\mu^2=M_{Pole}^2+\Gamma_{Pole}^2/4$ \cite{einhorn}.
Its position is also shown in the table. 
\begin{table}
\begin{center}
\begin{tabular}{|c|c|c|c|c|c|}
\hline
$m_h$    & $M_{Pole}$  & $\Gamma_{Pole}$     & $\Lambda_L$      & $\Lambda_T$       & $\kappa$ \\\hline
0.20   &  0.20  & 5.3$\times 10^{-3}$  & 1.7$\times 10^{25}$&1.7$\times 10^{25}$
&1.6$\times 10^{-50}$\\\hline
0.40   &  0.40  & 0.04   & 1.2$\times 10^6$ & 1.2$\times 10^6 $ & 0.1$\times 10^{-6}$ \\\hline
0.60   &  0.59  & 0.15   & 0.5$\times 10^3$ & 0.5$\times 10^3$  & 0.6$\times 10^{-3}$ \\\hline
0.80   &  0.74  & 0.34   & 3.2$\times 10 $  & 3.2$\times 10$    & 4.7$\times 10^{-3}$ \\\hline
1.00   &  0.83  & 0.58   & 9.64           & 9.88            & 4.7$\times 10^{-2}$ \\\hline
1.20   &  0.87  & 0.82   & 5.03           & 5.45            & 0.14              \\\hline
1.40   &  0.86  & 1.00   & 3.36           & 3.93            & 0.24              \\\hline
1.60   &  0.84  & 1.11   & 2.57           & 3.23            & 0.31              \\\hline
1.80   &  0.82  & 1.19   & 2.12           & 2.85            & 0.37              \\\hline
2.00   &  0.81  & 1.24   & 1.85           & 2.62            & 0.41        \\  \hline                    
\end{tabular}
\caption{\label{tachsize} {\em Information on the physical 
and the tachyonic pole at leading order.
The parameter $m_h$ is related to the physical pole in the
way explained in the text. 
The dimensionful quantities are in TeV.}}
\end{center}
\end{table} 
Near the Higgs resonance, the residuum of the tachyonic pole is 
a measure for the quantitative influence of 
the tachyonic regularization. Any amplitude  will 
be modified  there by terms proportional to $\kappa$. 
The perturbative expansion of these terms vanishes identically. 

Another strategy to treat the tachyon issue would 
be the introduction of a momentum cutoff \cite{schnitzer:cutoff}. 
Such a cutoff has to be under the tachyon scale, and
destroys the gauge invariance in the gauged model.
On the contrary, our tachyonic regularization preserves the symmetry
of the model in all orders of perturbation theory.

\subsection{Diagrammatics}

\begin{figure}
\begin{center}
\begin{picture}(13,4) 
 \put(+3.5,+0.){\epsfxsize6cm  \epsfysize3.75cm  \epsffile{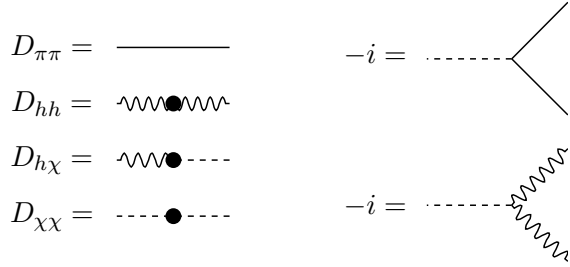}}
        \put(+2.7,+3.0){\makebox(0,0){$D_{\pi\pi} = $ }}
        \put(+2.7,+2.25){\makebox(0,0){$D_{hh} =$ }}
        \put(+2.7,+1.5){\makebox(0,0){$D_{h\chi} =$ }}
        \put(+2.7,+0.75){\makebox(0,0){$D_{\chi\chi} =$ }}
        \put(+7,+2.9){\makebox(0,0){$-i =$ }}
        \put(+7,+0.9){\makebox(0,0){$-i =$ }}
\end{picture}
\end{center}
\caption{\label{feynrules} {\em Feynman rules. 
The dot indicates the dressed, tachyon free propagators.
The counterterms are defined by eq. \ref{counterlag}.}}
\end{figure}

The Feynman rules for the next-to-leading order calculation contain the
full LO propagators given above in eq. \ref{propa} and the interaction terms 
defined in eqs. \ref{lagrenor}. The propagators are 
understood as tachyonicaly regulated by the prescription 
of eqs. \ref{tachsub}.
Thus the diagrammatical part of the calculation is settled. 
The Feynman rules are given in figure \ref{feynrules}.

The dot indicates the dressed propagators stemming from
the leading order calculation. We now fix the counterterms.
Considering the values of the leading order renormalization constants,
 we write $Z_\kappa = 1 - \kappa/N$, 
$Z_\phi= 1+z_\phi/N$, $Z_h= 1+z_h/N$, $Z_v^2 = 1 + 2 z_v/N$,
$V_\chi=v_\chi/N$, $1/Z_\lambda = 1 - z_\lambda/N$, $Z_\chi=1+z_\chi/N$.
The LO counterterm is already absorbed in the definition of the 
dressed propagators. The NLO counterterm Lagrangean 
is expressible in terms of these new variables:

\begin{eqnarray}\label{counterlag}
{\cal L}_{counterterm} &=& \frac{1}{N} \Biggl[
\frac{z_\phi}{2} \partial_{\nu} \vec\phi \partial^{\nu} \vec\phi +
Nv^2 \Bigl(\frac{v_\chi}{m_h^2} + z_v \Bigr) \chi +
 \frac{Nv^2}{m_h^2 } \left( 2 z_\chi-\kappa-z_\lambda\right) \frac{\chi^2}{2}
\nonumber \\ && \qquad
-\frac{z_\chi-\kappa}{2} \chi ( \vec\phi^2 - N v^2 ) 
-\frac{v_\chi}{2} ( \vec\phi^2 - N v^2 ) 
-\frac{m_h^2 \kappa}{8Nv^2} ( \vec\phi^2 - N v^2 )^2
  \Biggr]   \nonumber \\
\vec\phi &=& (\vec\pi,\sqrt{\frac{Z_h}{Z_\phi}}\sigma+\sqrt{N}v)
\end{eqnarray}

As we will see below, the NLO corrections are expressible in form of 
multi-loop diagrams made out of the dressed propagators. The counterterms
in eq. \ref{counterlag}
will allow to make subtractions of subgraphs inside these diagrams which
is the motivation for our redundant renormalization formalism.
 
The relevant two- and three-point functions are shown in figure \ref{qcor}.
Their actual form in terms of  graphs will be shown below. 

\begin{figure}
\begin{center}
\begin{tabular}{rcrcrc}
$\Gamma_{\chi\chi}=$ & \raisebox{-.4cm}{\epsfxsize2cm  \epsfysize1.cm  \epsffile{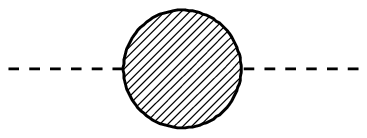}} &
$\Gamma_{hh}      =$ & \raisebox{-.4cm}{\epsfxsize2cm  \epsfysize1.cm  \epsffile{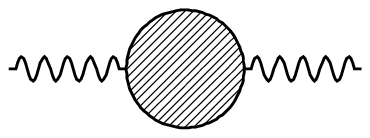}} & 
$\Gamma_{\chi h}  =$ & \raisebox{-.4cm}{\epsfxsize2cm  \epsfysize1.cm  \epsffile{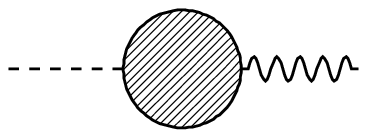}} \\
$\Gamma_h      =$ & \raisebox{-.4cm}{\epsfxsize2cm  \epsfysize1.cm  \epsffile{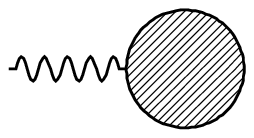}} &
$\Gamma_\chi         =$ & \raisebox{-.4cm}{\epsfxsize2cm  \epsfysize1.cm  \epsffile{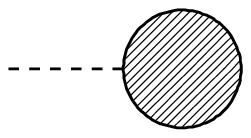}} & 
$\Gamma_{\pi\pi}  =$ & \raisebox{-.4cm}{\epsfxsize2cm  \epsfysize1.cm  \epsffile{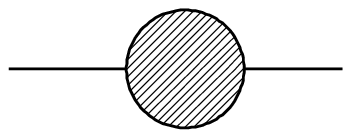}} \\ 
$\Gamma_{\chi\pi\pi}=$ & \raisebox{-.4cm}{\epsfxsize2cm  \epsfysize1.cm  \epsffile{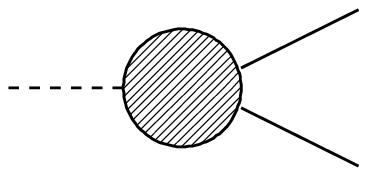}} &
$\Gamma_{h\pi\pi}=$ & \raisebox{-.4cm}{\epsfxsize2cm  \epsfysize1.cm  \epsffile{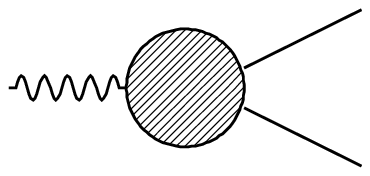}} 
\end{tabular}
\caption{\label{qcor} {\em Relevant two- and three-point functions 
                           to be calculated.}}
\end{center}
\end{figure}

The counterterms are fixed by some renormalization conditions.
Two conditions are already dictated by the 
cancellation of tadpole contributions.
Further conditions may be imposed on one-particle irreducible 
two- and three-point functions. 
We consider the following conditions:
 
\begin{eqnarray}\label{renscheme} 
\Gamma_{\pi\pi}'(0)=\Gamma_{hh}'(0)=
\frac{m_h^2}{Nv^2}\Gamma_{\chi\chi}(\mu^2)=\frac{-1}{\sqrt{N}v}\Gamma_{h\chi}=1, \quad
\Gamma_\chi = \Gamma_\sigma = \Gamma_{h\pi\pi}(0,0) = 0
\end{eqnarray}

These conditions define a renormalization scheme. 
Otherwise, all possible counterterms
are fixed by the symmetry of the counterterm Lagrangean of 
eq. \ref{counterlag}. 
Whenever possible we set the renormalization scale to zero,
because the multiloop counterterms turn out to be analytically simpler,
as will become clear later. 
When the subtraction point $\mu$ cannot be set to zero because
of infrared problems, we combined the multiloop graph with
the corresponding counterterms at an Euclidian $\mu$ and
such that the sum is free of the infrared singularity,
and afterwards we took the limit $\mu \rightarrow 0$.

\begin{figure}
\begin{tabular}{llllll}
$\tilde A_1=$ &   \raisebox{-.6cm}{\epsfxsize3cm  \epsfysize1.5cm  \epsffile{a1.eps}}&
$\tilde A_2=$ &   \raisebox{-.6cm}{\epsfxsize3cm  \epsfysize1.5cm  \epsffile{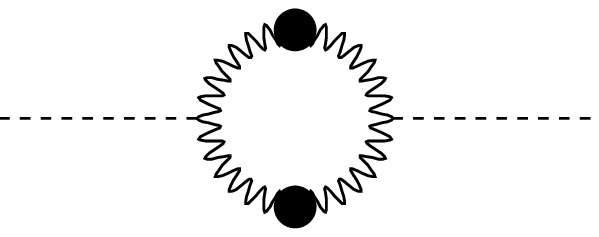}}&
$\tilde A_3=$ &   \raisebox{-.6cm}{\epsfxsize3cm  \epsfysize1.5cm  \epsffile{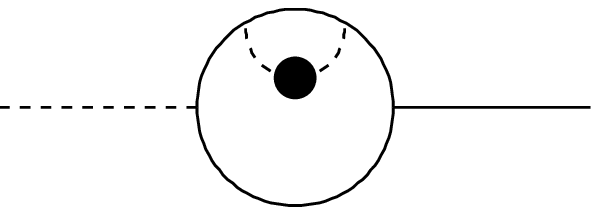}}\\
$\tilde A_4=$ &   \raisebox{-.6cm}{\epsfxsize3cm  \epsfysize1.5cm  \epsffile{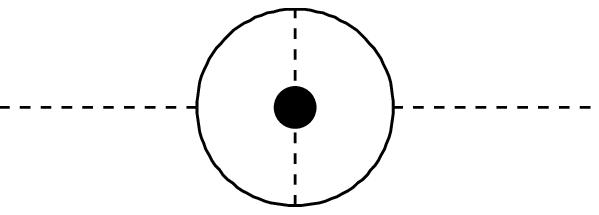}}&
$\tilde A_5=$ &   \raisebox{-.6cm}{\epsfxsize3cm  \epsfysize1.5cm  \epsffile{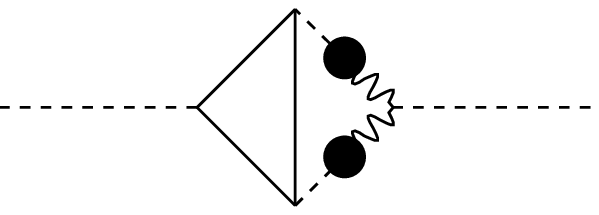}}&
$\tilde A_6=$ &   \raisebox{-.6cm}{\epsfxsize3cm  \epsfysize1.5cm  \epsffile{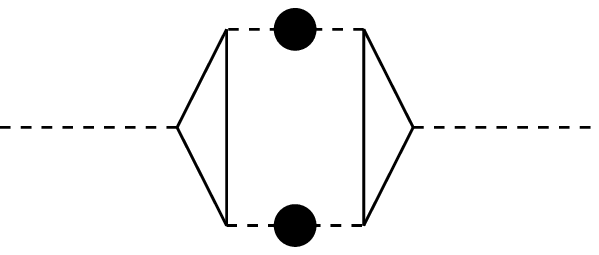}}\\
$\tilde A_7=$ &   \raisebox{-.6cm}{\epsfxsize3cm  \epsfysize1.5cm  \epsffile{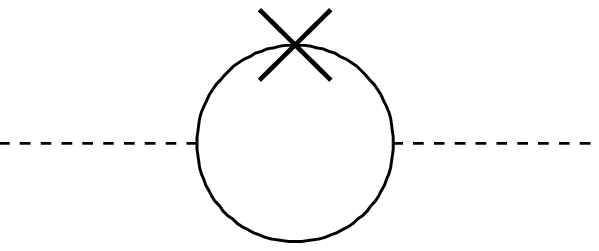}}&
$\tilde A_8=$ &   \raisebox{-.6cm}{\epsfxsize3cm  \epsfysize1.5cm  \epsffile{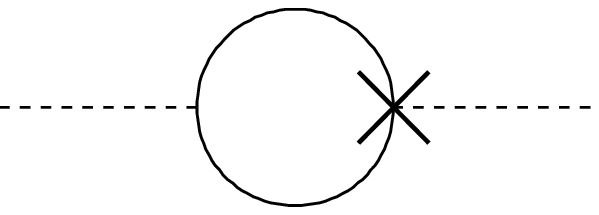}}&
$\tilde A_9=$ &   \raisebox{-.6cm}{\epsfxsize3cm  \epsfysize1.5cm  \epsffile{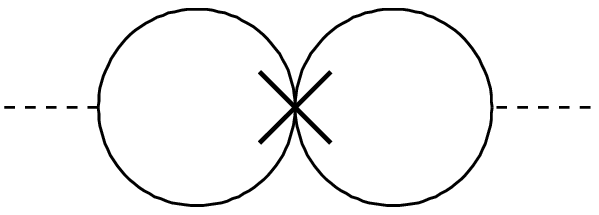}}\\
$\tilde B_1=$ &   \raisebox{-.6cm}{\epsfxsize3cm  \epsfysize1.5cm  \epsffile{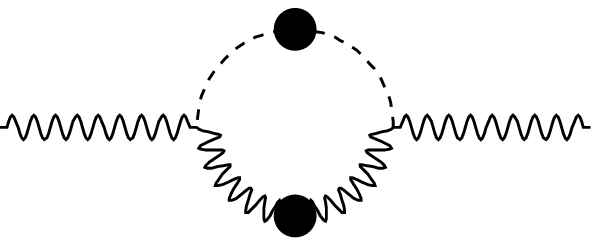}}&
$\tilde B_2=$ &   \raisebox{-.6cm}{\epsfxsize3cm  \epsfysize1.5cm  \epsffile{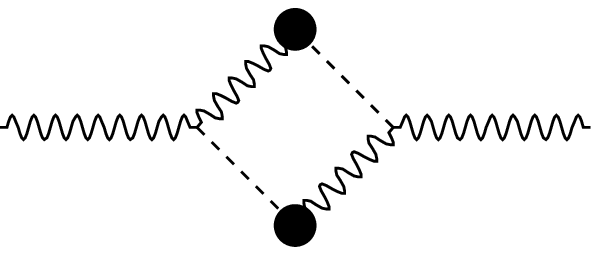}}&
$\tilde C_1=$ &   \raisebox{-.6cm}{\epsfxsize3cm  \epsfysize1.5cm  \epsffile{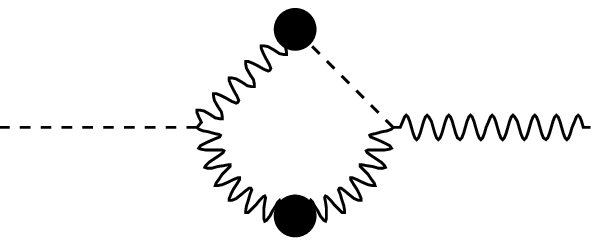}}\\
$\tilde C_2=$ &   \raisebox{-.6cm}{\epsfxsize3cm  \epsfysize1.5cm  \epsffile{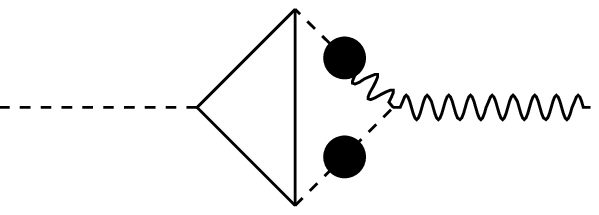}}&
$\tilde C_3=$ &   \raisebox{-.6cm}{\epsfxsize3cm  \epsfysize1.5cm  \epsffile{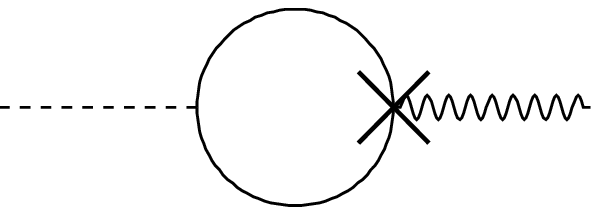}}&
$\tilde D=  $ &   \raisebox{-.6cm}{\epsfxsize3cm  \epsfysize1.5cm  \epsffile{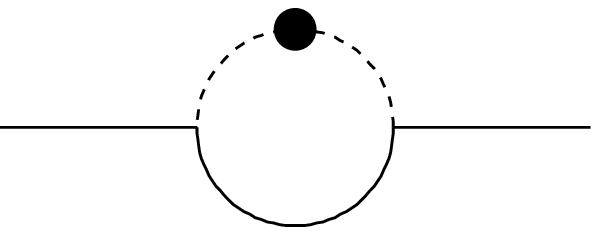}}\\
$\tilde E  =$ &   \raisebox{-.6cm}{\epsfxsize3cm  \epsfysize1.5cm  \epsffile{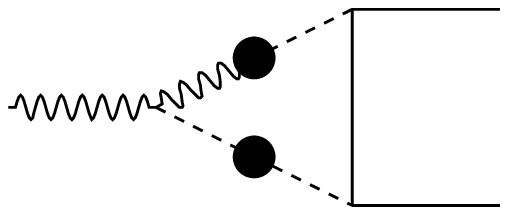}}&
$\tilde F_1=$ &   \raisebox{-.6cm}{\epsfxsize3cm  \epsfysize1.5cm  \epsffile{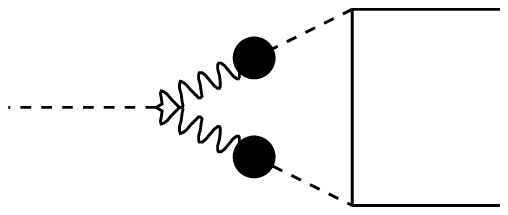}}&
$\tilde F_2=$ &   \raisebox{-.6cm}{\epsfxsize3cm  \epsfysize1.5cm  \epsffile{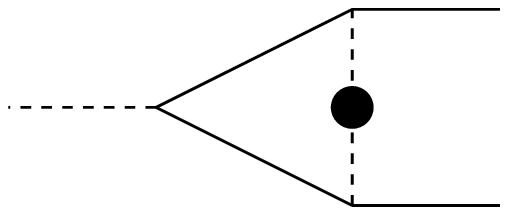}}\\
$\tilde F_3=$ &   \raisebox{-.6cm}{\epsfxsize3cm  \epsfysize1.5cm  \epsffile{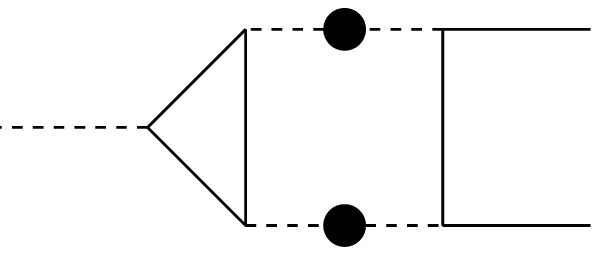}}&
$\tilde F_4=$ &   \raisebox{-.6cm}{\epsfxsize3cm  \epsfysize1.5cm  \epsffile{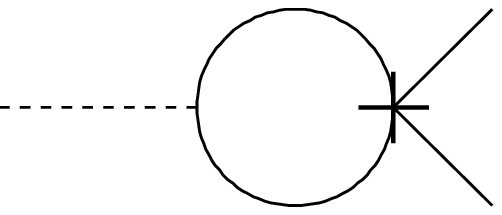}}&
$\tilde T_h=$ &   \raisebox{-.6cm}{\epsfxsize3cm  \epsfysize1.5cm  \epsffile{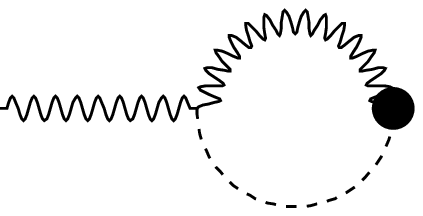}}
\end{tabular}
\caption{\label{1ongraphs} {\em List of occurring graphs to order $1/N$ }}
\end{figure}

The list of needed graphs is given in figure \ref{1ongraphs}. 
One finds these graphs by usual power counting in $1/N$.
A detailed discussion of their computation will follow in the 
next subsection where the 
analytical part of the calculation is done.
The renormalization constants can be determined 
through the scheme of eq. \ref{renscheme}
in terms of vacuum graphs of the type given in figure \ref{1ongraphs}. 
By using 

\begin{eqnarray}
\tilde A(\mu^2)&=&-\tilde A_1(-\mu^2)/2+\tilde A_2(0)/2-\tilde A_3(0)-
\tilde A_4(0)/2-\tilde A_5(0)+\tilde A_6(0)/2\nonumber\\
 && +\tilde A_7(-\mu^2)-\tilde A_8(-\mu^2)+\tilde A_9(-\mu^2)/4 \nonumber\\
\tilde C(\mu^2)&=&\tilde C_1(0)-\tilde C_2(0)+\tilde C_3(-\mu^2)\nonumber
\end{eqnarray}
one finds

\begin{eqnarray}\label{fixedrencon}
\begin{array}{lll}
z_\phi = -v^2 \tilde D'(0) & z_h = -v^2 \tilde B_1'(0) &
z_v    = -\tilde T_\chi - \frac{v^2 \tilde T_h}{m_h^2} \\
z_\chi = \frac{2v^2}{m_h^2}\tilde B_2(0) + \tilde C(0) &
\kappa = \frac{2v^2}{m_h^2}\tilde B_2(0) &
v_\chi = v^2 \tilde T_h \\
z_\lambda = \frac{m_h^2}{v^2} \tilde A(\mu^2) + \frac{2v^2}{m_h^2}\tilde B_2(0) +2 \tilde C(\mu^2)&&
\end{array}
\end{eqnarray}

Note that the occurring expressions can be expanded in $\lambda$. This 
provides the connection to the standard loop expansion, 
and allows to check the consistency of the 
renormalization procedure to any order diagramatically.

\subsection{Analytical framework}

In order to keep our expressions compact, we introduce the 
following notations:
  
\begin{equation}
g(s) = \frac{1}{s-m^2(s)}\, , \quad f(s) = \frac{m^2(s)}{s-m^2(s)}
\end{equation} 
We further do not write down explicitly the tachyonic regularization
of the propagators, but we understand that the tachyonic pole has
to be subtracted according to the discussion of the previous section
before numerical integration is performed.

To obtain compact formulae for the occurring $1/N$ 
graphs, we use in the following the massless scalar one-loop
three- and four-point functions as building blocks.  
Their explicit expressions \cite{ghinculov:3loop,denner:box}
are listed in the appendix.
With the help of these functions, the two- and three-loop topologies 
can be cast into the form of two-fold integrals.
The $1/N$ graphs which we need correspond to the 
following list of dimensionless integrals,
where $r=(q+k)$ and we use the notation $dQ=\frac{d^4q}{(2\pi)^4i}$.
$k$ and $\mu$ stand for a timelike and spacelike four-vector respectively, 
and $l_\mu$ is a lightlike four-vector corresponding to on-shell GBs. 
To work with dimensionless quantities, we divide the $1/N$ graphs 
by appropriate factors of $v$.
We distinguish ultraviolet subtracted graphs from 
unsubtracted ones by a tilde.

\subsubsection{One-loop topologies}

\begin{eqnarray}
\tilde T_h  &=& \frac{1}{v^4}  \int dQ f(q^2)\\
A_1  &=&  -\int dQ \frac{1}{q^2}\left(\frac{1}{r^2}-\frac{1}{(q+\mu)^2}\right)
            =\frac{1}{(4\pi)^2}\log\Bigl(\frac{-k^2}{\mu^2}\Bigr)\\
A_2   &=&   \int dQ g(q^2)\Bigl[g(r^2)-g(q^2)\Bigr] \\
B_1   &=&   \frac{1}{v^4}  \int dQ q^2f(q^2)\Bigl[ g(r^2)-g(q^2)-
                k^2\frac{d}{dk^2}\vert_{k^2=0}\frac{1}{r^2}\Bigr] \\
B_2   &=&   \frac{1}{v^2}  \int dQ  f(q^2)\Bigl(f(r^2)-f(q^2)\Bigr)\\ 
C_1   &=& \frac{1}{v^2} \int dQ f(q^2)\Bigl[g(r^2)-g(q^2)\Bigr] \\
D    &=& \frac{1}{v^4}  \int dQ q^2 f(q^2) \left( \frac{1}{r^2} -\frac{1}{q^2} -  
                k^2\frac{d}{dk^2}\vert_{k^2=0}\frac{1}{r^2}\right)\\
E &=& \frac{1}{v^2}  \int dQ \left[ \frac{q^2}{(q+l)^2} f(q^2)f((q-k)^2) - f^2(q^2)\right] \\
F_1 &=&\frac{1}{v^2}  \int dQ f(q^2) \left[ \frac{q^2}{(q-l)^2(q-l+k)^2}-\frac{1}{q^2}\right] \\
F_2 &=&\frac{1}{v^2}  \int dQ \left[\frac{q^2}{(q+l)^2} f(q^2)f((q-k)^2) - f^2(q^2)\right] \\
\end{eqnarray}

All necessary counterterms for the subtractions
appear in the counter Lagrangean of eq.~(\ref{counterlag}), 
as one can show by using the relations 
$\tilde T_h=\tilde D(0)=\tilde B_1(0)-\tilde B_2(0)$, 
$\tilde F_1(0)+\tilde F_2(0)=\tilde C_1(0)$.
For convenience we define
$B_1(s)=\tilde B_1(s)-\tilde B_1(0)-s \tilde D'(0)$ because we 
will not explicitly  need the  
wave function renormalization of the Higgs boson 
which is related to $\tilde B'$. 
Note that $\tilde T_h=\tilde D(0)$ is a  necessary consistency condition 
to keep the 
GBs massless after summing the perturbative series. This holds
independently of the tachyon. 
One can check that to render $\tilde B_1$ and $\tilde D$ finite, one subtraction is
already sufficient. This means that one gets an ultraviolet convergent 
wave function renormalization constant, 
in contrast to usual perturbation theory.  
Also note that graphs like $\tilde B_2$, 
with two momentum dependent masses in the
numerator, have an extra suppression factor of $1/log^2(q^2)$ 
in the ultraviolet,  
which leads to a convergent 
integral\footnote{The integral converges as $1/\log{\Lambda_{cutoff}}$}.
The summation of all loop graphs thus improves 
the UV behaviour.  
This ultraviolet screening is a remarkable feature of the
nonperturbative $1/N$ expansion.  
   
\subsubsection{Higher-loop topologies}

\begin{figure}
\begin{tabular}{rrrr}
$D=\quad$ \raisebox{-.5cm}{\epsfxsize2.4cm \epsfysize1.2cm \epsffile{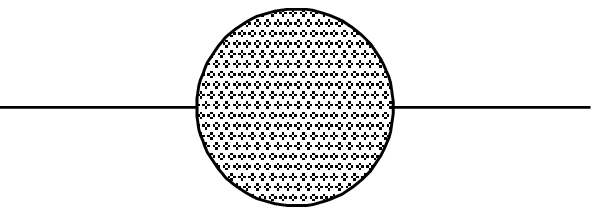}} &
$=\quad$  \raisebox{-.5cm}{\epsfxsize2.4cm \epsfysize1.2cm \epsffile{d.eps}} & 
$-\quad$  \raisebox{-.5cm}{\epsfxsize2.4cm \epsfysize1.2cm \epsffile{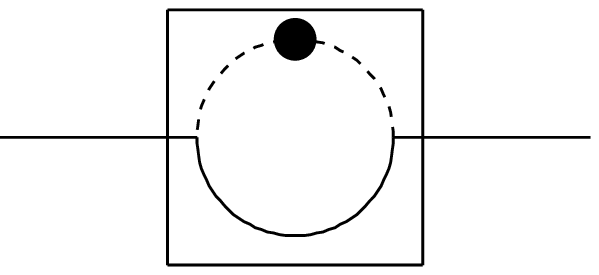}} & 
$-s$      \raisebox{-.5cm}{\epsfxsize2.4cm \epsfysize1.2cm \epsffile{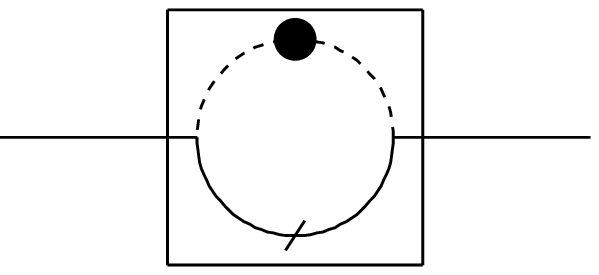}} \\  
$A_3=\quad$ \raisebox{-.5cm}{\epsfxsize2.4cm\epsfysize1.2cm \epsffile{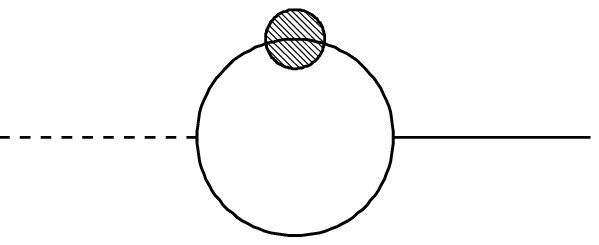}} & 
$-\quad$   \raisebox{-.5cm}{\epsfxsize2.4cm\epsfysize1.2cm \epsffile{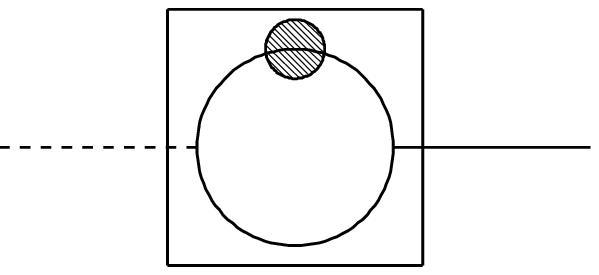}}& &  \\
$A_4=\quad$ \raisebox{-.5cm}{\epsfxsize2.4cm\epsfysize1.2cm \epsffile{a4.eps}} & 
$-2\quad$   \raisebox{-.5cm}{\epsfxsize2.4cm\epsfysize1.2cm \epsffile{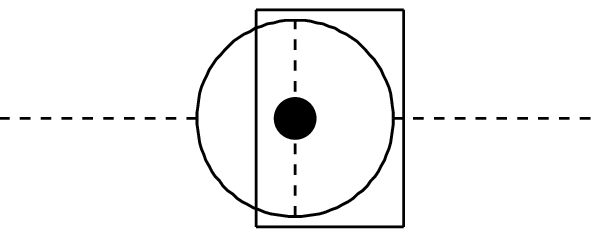}}&
$-\quad$   \raisebox{-.5cm}{\epsfxsize2.4cm\epsfysize1.2cm \epsffile{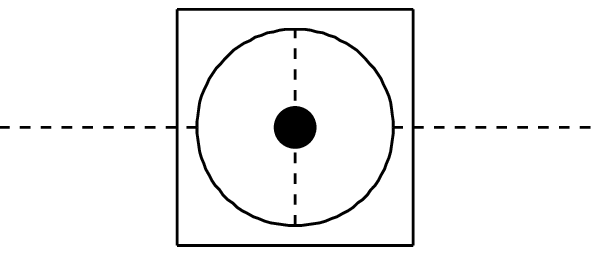}} &
$+2\quad$   \raisebox{-.5cm}{\epsfxsize2.4cm\epsfysize1.2cm \epsffile{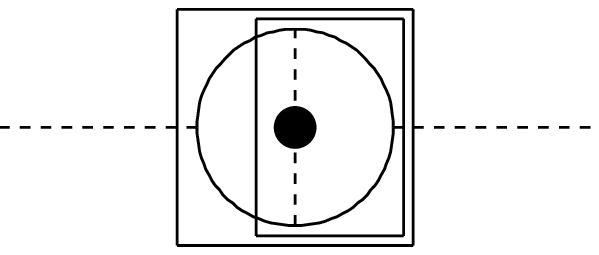}} \\
$A_5=\quad$ \raisebox{-.5cm}{\epsfxsize2.4cm\epsfysize1.2cm \epsffile{a5.eps}}  & 
$-\quad$   \raisebox{-.5cm}{\epsfxsize2.4cm\epsfysize1.2cm \epsffile{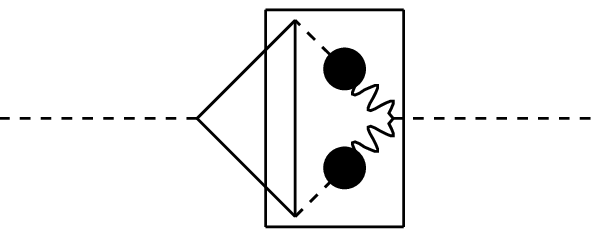}} &
$-\quad$   \raisebox{-.5cm}{\epsfxsize2.4cm\epsfysize1.2cm \epsffile{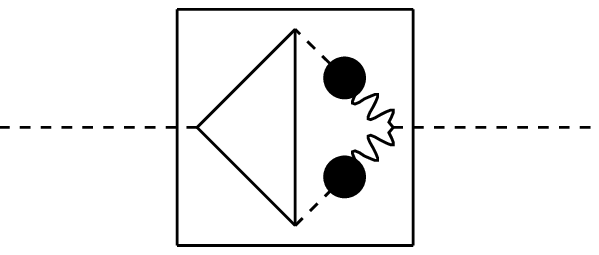}} &
$+\quad$   \raisebox{-.5cm}{\epsfxsize2.4cm\epsfysize1.2cm \epsffile{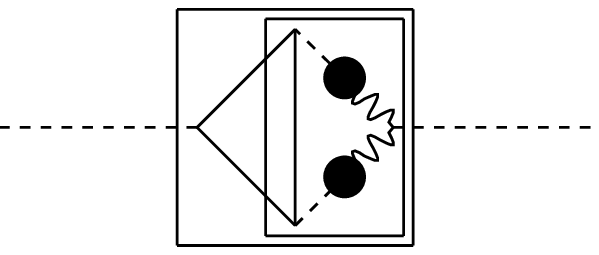}} \\
$A_6=\quad$ \raisebox{-.5cm}{\epsfxsize2.4cm\epsfysize1.2cm \epsffile{a6.eps}} & 
$-\quad$   \raisebox{-.5cm}{\epsfxsize2.4cm\epsfysize1.2cm \epsffile{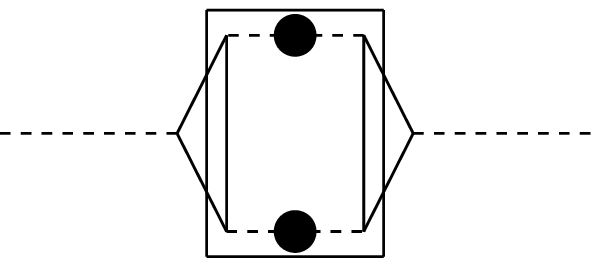}}&
$-2$   \raisebox{-.5cm}{\epsfxsize2.4cm\epsfysize1.2cm \epsffile{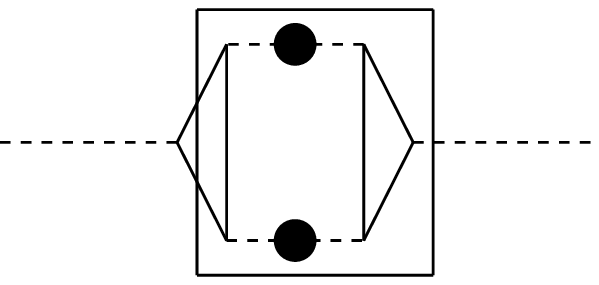}} &
$+2$   \raisebox{-.5cm}{\epsfxsize2.4cm\epsfysize1.2cm \epsffile{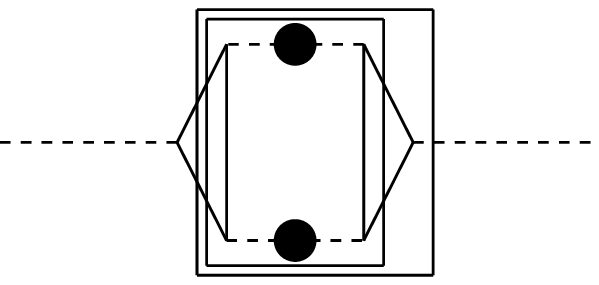}}\\
     &&& \\
& $-\,$ overall subtr.  & &  \\
     &&& \\ 
$C_2=\quad$ \raisebox{-.5cm}{\epsfxsize2.4cm\epsfysize1.2cm \epsffile{c2.eps}} & 
$-\quad$   \raisebox{-.5cm}{\epsfxsize2.4cm\epsfysize1.2cm \epsffile{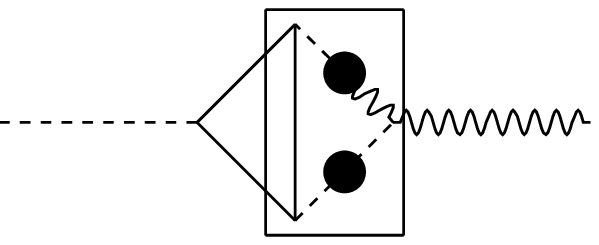}}&
$-\quad$   \raisebox{-.5cm}{\epsfxsize2.4cm\epsfysize1.2cm \epsffile{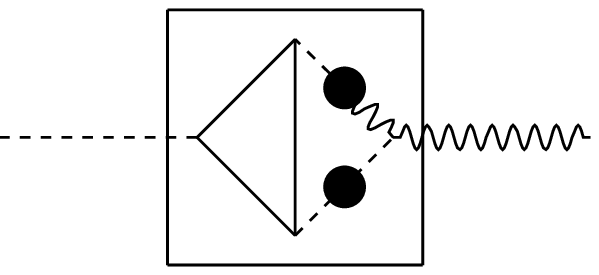}} &
$+\quad$   \raisebox{-.5cm}{\epsfxsize2.4cm\epsfysize1.2cm \epsffile{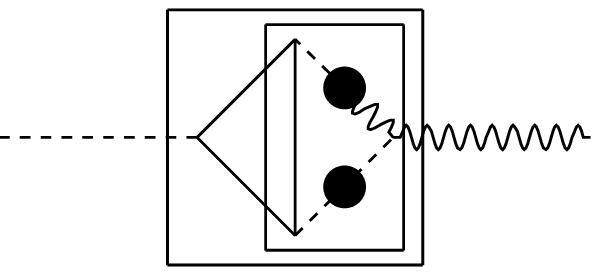}} \\
$F_3=\quad$ \raisebox{-.5cm}{\epsfxsize2.4cm\epsfysize1.2cm \epsffile{vxpp3.eps}} & 
$-\quad$   \raisebox{-.5cm}{\epsfxsize2.4cm\epsfysize1.2cm \epsffile{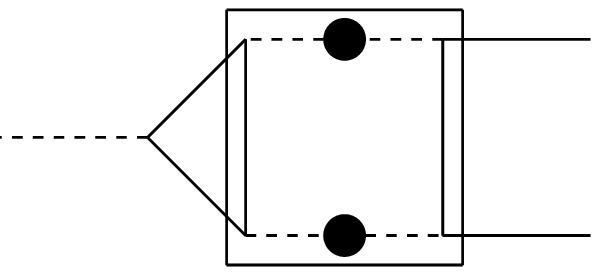}} &
$-\quad$   \raisebox{-.5cm}{\epsfxsize2.4cm\epsfysize1.2cm \epsffile{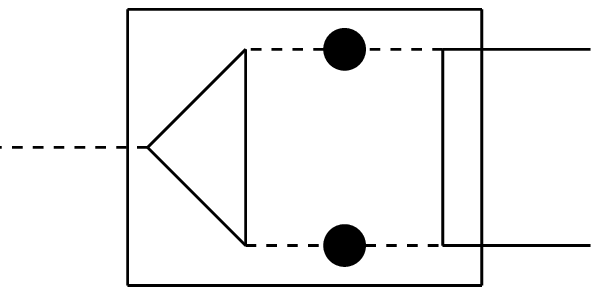}} &
$+\quad$   \raisebox{-.5cm}{\epsfxsize2.4cm\epsfysize1.2cm \epsffile{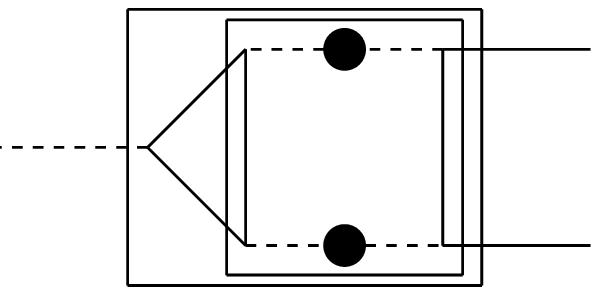}}
\end{tabular}
\caption{\label{graphisub} {\em Subtraction scheme of the higher loop topologies.
In the countergraph of D, the dash stands for differentiation with respect to the external 
momentum and the box means that the contained subgraphs 
have to be evaluated at vanishing external momenta.}}
\end{figure}

For the higher-loop topologies one has to subtract the 
subgraphs first. The necessary subtractions are most transparent in a 
graphical representation. We have collected them in figure \ref{graphisub}. 

\begin{figure}
\begin{center}
\begin{tabular}{rrr}
\raisebox{-.4cm}{\epsfxsize2.cm \epsfysize1.cm \epsffile{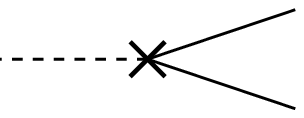}} & 
$=\quad$  \raisebox{-.6cm}{\epsfxsize2.8cm \epsfysize1.4cm \epsffile{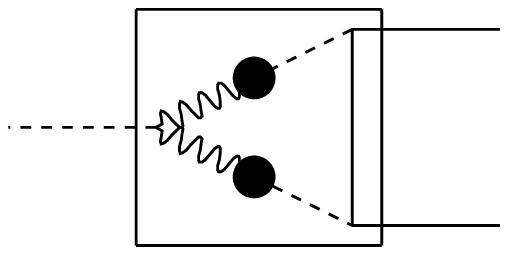}} & 
$+\quad$  \raisebox{-.6cm}{\epsfxsize2.8cm \epsfysize1.4cm \epsffile{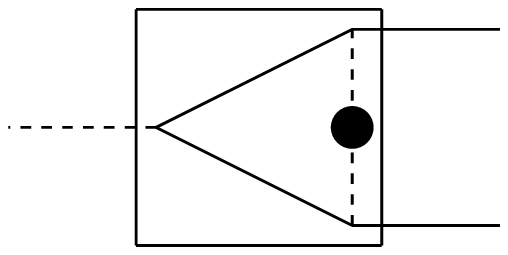}} \\ &
$-\quad$  \raisebox{-.6cm}{\epsfxsize2.8cm \epsfysize1.4cm \epsffile{vxpp3c2.eps}}  &  
$+\quad$  \raisebox{-.6cm}{\epsfxsize2.8cm \epsfysize1.4cm \epsffile{vxpp3c3.eps}} 
%
\end{tabular}
\end{center}
\caption{\label{3counter} {\em Graphical representation of the 
$\Gamma_{\chi\pi\pi}$ counterterm.}}
\end{figure}

The reader can check that the counterterms defined in eq. \ref{counterlag}
correspond to sums of boxed graphs. For the three-point counterterms,
this is shown in figure \ref{3counter}.

Note that the sum of the $1/N$ graphs and counterterms 
for a given topology gives an infrared finite result. Thus, to derive
the following formulae, it is necessary to introduce an infrared 
regulator for the individual terms
which cancels in the final sum. As stated above, we are 
using analytical expressions for the massless three- and four- point
functions, $C_0$, $D_0$, which are given in an appendix for completeness.

\begin{eqnarray}\label{integralreps}
A_3   &=& \frac{1}{v^2} \int dQdP 
       \frac{f(q^2) q^2}{(p^2)^2}\left[\frac{1}{(p+k)^2}-\frac{1}{q^2}\right]
      \left[\frac{1}{(p+q)^2}-\frac{1}{p^2}-\frac{q^2}{(p^2)^2}
      \Bigl(1-\frac{4(p\cdot q)^2}{p^2q^2}\Bigr)\right]  \nonumber\\
   &=&  \frac{1}{v^2} \int dQ  \frac{f(q^2)}{q^2 (4\pi)^2}
   \left[ 1-\frac{2(q\cdot k)^2}{q^2k^2} -\frac{q^2}{k^2}\log\Bigl(\frac{r^2}{q^2}\Bigr) \right]
\nonumber\\
A_4 &=&  \frac{1}{v^2} \int dQ  \frac{f(q^2)}{q^2}
       \left] (q^2)^2 D_0(k,q,-k) - \frac{1+\log\bigl(\frac{q^2}{k^2}\bigr)}{8\pi^2} \right]
\nonumber\\
A_5 &=&  \frac{1}{v^2} \int dQ  
       \left[ f(r^2)f(q^2) C_0(k,q) - 
   f^2(q^2)  \frac{2+\log\bigl(\frac{q^2}{k^2}\bigr)}{(4\pi)^2q^2} \right]
\nonumber\\
A_6 &=&  \frac{1}{v^4} \int dQ  
       \left[ r^2 q^2f(r^2)f(q^2) C_0^2(k,q) - 
   f^2(q^2)  \left(\frac{2+\log\bigl(\frac{q^2}{k^2}\bigr)}{(4\pi)^2}\right)^2 \right]
\nonumber\\
C_2 &=&  \frac{1}{v^2} \int dQ  
       \left[ q^2 f(r^2)f(q^2) C_0(k,q) - 
   f^2(q^2)  \frac{2+\log\bigl(\frac{q^2}{k^2}\bigr)}{(4\pi)^2} \right]
\nonumber\\
F_3 &=&  \frac{1}{v^4} \int dQ  
       \left[ f(r^2)f(q^2) C_0(k,q) \frac{r^2 q^2}{(q-l)^2} - 
   f^2(q^2)  \frac{2+\log\bigl(\frac{q^2}{k^2}\bigr)}{(4\pi)^2} \right]
\end{eqnarray}
We will need in the following sums of these  expressions 
taking into account combinatorial factors and signs.
\begin{eqnarray}\label{rengammas}
A(s) &=& -\frac{1}{2} A_1(s) + \frac{1}{2} A_2(s) -
          A_3(s) - \frac{1}{2} A_4(s) - A_5(s) +  \frac{1}{2} A_6(s) \nonumber\\
B(s) &=& B_1(s) + B_2(s) \nonumber \\
C(s) &=& C_1(s) - C_2(s) \nonumber\\ 
F(s) &=& F_1(s) + F_2(s) - F_3(s) 
\end{eqnarray}
The functions $A,B,C,D,E,F$ are integrals over Minkowski space, 
parameterized by 
$q_\nu=(q_0,\rho \vec n)$.
After performing the angular integration, 
the number of integrations can be reduced to two.
Choosing the timelike momentum as $k_\nu=(\sqrt{s},\vec 0)$, it is obvious that the
integrals which belong to two-point functions  
are isotropic. For the three-point function with on-shell GBs, 
one has a second timelike 
four-vector $l_\nu=(1,0,0,1) \sqrt{s}/2$.
Explicitly, one finds for  the angular integration:

\begin{eqnarray}\label{angint}
\int \frac{d\Omega}{4\pi} \frac{1}{(q-l)^2} &=& 
 \frac{1}{2\rho\sqrt{s}}\log\left[\frac{q^2+\sqrt{s}(q_0+\rho)}{q^2+\sqrt{s}(q_0-\rho)}\right]
\nonumber\\
\int \frac{d\Omega}{4\pi}  \frac{1}{(q-l)^2(q-l+k)^2} &=& 
\frac{1}{4q_0\rho s}\log\left[\frac{(q^2)^2-s (q_0-\rho)^2}{(q^2)^2-s (q_0+\rho)^2}\right]
\end{eqnarray}
Thus we end up with two-dimensional integral representations for the relevant
graphs contributing to the processes we look for. In principle
the $q_0$ integration is also manageable by complex rotation, but one
has to take great care because of the complicated singularity and cut
structure of the integrands. We proceed numerically in the following. 

\subsection{Numerical calculation of the diagrams}

\begin{figure}
\begin{center}
\begin{picture}(10,3)
 \put(1,0.0){\epsfxsize4cm \epsfysize3cm \epsffile{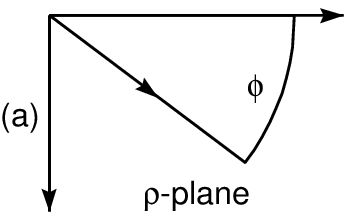}}
 \put(6,0.0){\epsfxsize4cm \epsfysize3cm \epsffile{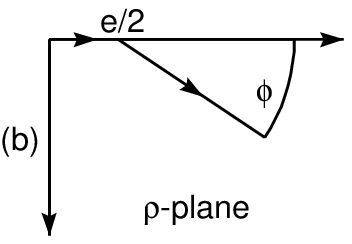}}
\end{picture}
\end{center}
\caption{\label{contours} {\em Integration contours for the numerical integration
of the (a) two- and (b) three-point functions. $e=\sqrt{s}$.}}
\end{figure}

The numerical problems encountered in the calculation of 
the needed diagrams are twofold.
First of all, one faces convergence problems due to the physical particle
poles. Although the massive propagators have a finite width in our case,
one still has to use a  Wick rotation to get a fast numerical convergence.
To do the two dimensional numerical integration:

\begin{eqnarray}
\int\limits_{-\infty}^{\infty} dq_0 \int\limits_{0}^{\infty}d\rho \rho^2 f(q_0,\rho) \nonumber
\end{eqnarray}
one chooses the contours in the complex $\rho$ plane to 
avoid the cuts of the logarithms and to stay away from the poles.
They are shown in figure \ref{contours}.
For the two-point functions, a simple rotation $\rho\rightarrow \rho\exp(-i\phi)$
leads to a smooth integrand (a), for the three point function
one has to avoid also the cut of the logarithms 
present in  eq. \ref{angint} (b). 
No singularity and cut is crossed for any $\phi\in (0,\pi/2)$.
The independence of the numerical result of $\phi$ serves as a test
of the stability of the integration.

The procedure gives fast and accurate results as long as the integrand is well behaved 
at infinity.
This is the second problem of the numerical
part of the calculation. Usually in loop calculations one is able to separate
the UV divergent part by evaluating the Feynman graphs in
$4+\epsilon$ dimensions. The regularization leads to manifestly finite integrands.
If, as in our calculation, the regularization has to be done
by subtracting divergent counterterms from divergent graphs, 
the numerical situation
is more delicate because of cancellations. 
To ensure numerical ultraviolet stability, it is necessary 
to arrange the integrand in such a way that as many cancellations as possible
are manifest. For the integrations we used adaptive, deterministic
Fortran routines.

\section{The results}

To derive physical amplitudes, we collect our results for the 
two- and three- point functions.  
For the renormalized two-point functions one finds the following
expressions:

\begin{eqnarray}
\Gamma_{\{h,\chi\}}(s) &=&  \left[\begin{array}{cc} 
\Gamma_{\chi\chi} & \Gamma_{h\chi} \\
 \Gamma_{h\chi}   & \Gamma_{hh}   \end{array}\right]
= \left[\begin{array}{cc} 
\frac{Nv^2}{m^2(s)} + A(s) & 
     -\sqrt{N}v\left(1-\frac{C(s)}{N}\right)\sqrt{\frac{Z_h}{Z_\phi}}  \\
-\sqrt{N}v\left(1-\frac{C(s)}{N}\right)\sqrt{\frac{Z_h}{Z_\phi}} & 
\left( s + \frac{v^2 B(s)}{N}\right)\frac{Z_h}{Z_\phi} \end{array}\right] \nonumber \\
\Gamma_{\pi_k\pi_l}(s) &=&   s  \left[ 1 +\frac{v^2}{N} D(s) \right] \delta_{kl}
\end{eqnarray}

The leading order contribution is contained in the momentum dependent mass. 
The GBs, labeled by $k,l\in\{1,\dots,N-1\}$, do not mix with the 
$\chi,h$ fields.
The wave function renormalization constants appear because of our  definition
of $B_1$. As mentioned above, we used only the wave function renormalization of the
GBs, because we are interested in amplitudes without external Higgs particles. Thus
it is not necessary to compute $Z_h$. With the methods described in
the previous section, 
it is easy to calculate $Z_{\phi}-Z_h=\frac{B_1'-D'}{N}$.
The inverse of the two point function matrix
gives the propagators. As explained earlier, the diagonal GB propagators
still have their pole at $s=0$.
One finds:

\begin{eqnarray}
D_{hh}(s) &=& \frac{Z_\phi}{Z_h}\, \frac{1}{s - M^2(s)} \\
\mbox{with} \qquad \quad && \nonumber \\
\frac{1}{M^2(s)} &=& \frac{1}{m^2(s)} \left\{ 1 + \frac{1}{N} 
\Bigl[\frac{m^2(s)}{v^2} A(s)+2 C(s) + 
\frac{v^2}{m^2(s)} B(s)\Bigr] \right\}  \nonumber
\end{eqnarray}

The other propagators are:
\begin{eqnarray}
\frac{D_{h\chi}(s)}{D_{h\chi}^0(s)} &=& 
        \sqrt{\frac{Z_h}{Z_\phi}} \frac{D_{hh}(s)}{D_{hh}^0(s)} - 
   \frac{1}{N}\Bigl[\frac{m^2(s)}{v^2}A(s)+C(s)\Bigr]  \nonumber \\
\frac{D_{\chi\chi}(s)}{D_{\chi\chi}^0(s)} &=& 
     \frac{Z_h}{Z_\phi}   \frac{D_{hh}(s)}{D_{hh}^0(s)} - 
   \frac{1}{N}\Bigl[\frac{m^2(s)}{v^2}A(s)-\frac{v^2}{s}B(s)\Bigr] 
\end{eqnarray}
The three-point functions are:
\begin{eqnarray}
\Gamma_{h\pi\pi}(s) &=& - \frac{2 E(s)}{N}   \nonumber \\
\Gamma_{\chi\pi\pi}(s) &=& - 1 - \frac{ F(s) }{N}  
\end{eqnarray}

\begin{figure}
\begin{center}
\begin{picture}(14,2.2)
\put(0.1,0.0){\epsfxsize4cm \epsfysize2cm \epsffile{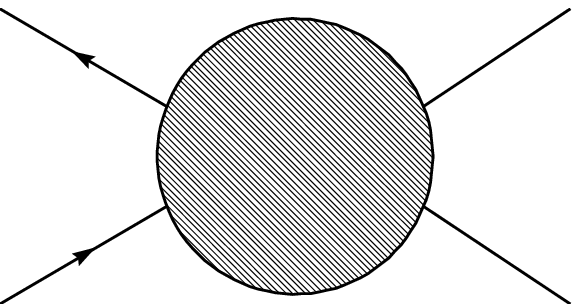}} 
\put(4.,.9){$=$}
\put(4.7,0.0){\epsfxsize4cm \epsfysize2cm \epsffile{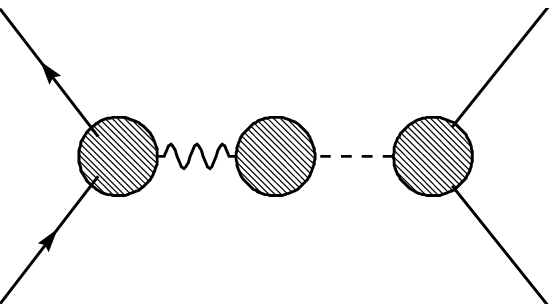}} 
\put(8.7,.9){$+$}
\put(9.4,0.0){\epsfxsize4cm \epsfysize2cm \epsffile{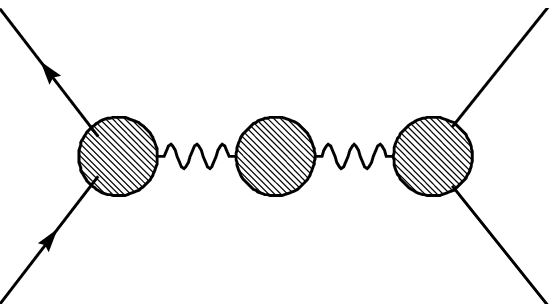}} 
%
\end{picture}
\end{center}
\caption{\label{mffzzamp} {\em Decomposition of the 
NLO 1/N corrections for the amplitude ${\cal M}(f\bar f\rightarrow zz)$ into
irreducible vertex corrections and dressed propagators.}}
\end{figure}

With these ingredients, one can express the amplitudes of the 
reactions $f\bar f\rightarrow f'\bar f'$ and   $f\bar f\rightarrow zz$.
Note that for any $Hf\bar f$ 
vertex one gets a multiplicative factor $\sqrt{Z_h/Z_\phi}$.
The decomposition of the amplitudes into one-particle 
irreducible contributions is  depicted in figure \ref{mffzzamp}.

\begin{eqnarray}
{\cal M}(f\bar f\rightarrow f'\bar f') &=& \frac{1}{s-M^2(s)} \\
{\cal M}(f\bar f\rightarrow z\,z\,) \,        &=& \frac{1}{s-M^2(s)} 
\frac{m^2(s)}{\sqrt{N}v} \nonumber \\ && \quad  \times
\left\{1-\frac{1}{N}\Bigl[\frac{m^2(s)}{v^2}A(s)+C(s)+\frac{2v^2}{m^2(s)}E(s)+F(s)\Bigr]\right\} 
\end{eqnarray}

The amplitudes can be measured directly at a future muon collider,
and also can be related to Higgs production by gluon fusion at the
LHC. The processes are available also to 
NNLO in usual perturbation theory,
and thus are  good quantities to compare the different theoretical approaches.
One has simply to put $N=4$ and $v=123$ GeV in our formulae.

\begin{figure}
\hspace{.5cm}
\epsfxsize = 12.5cm
\epsffile{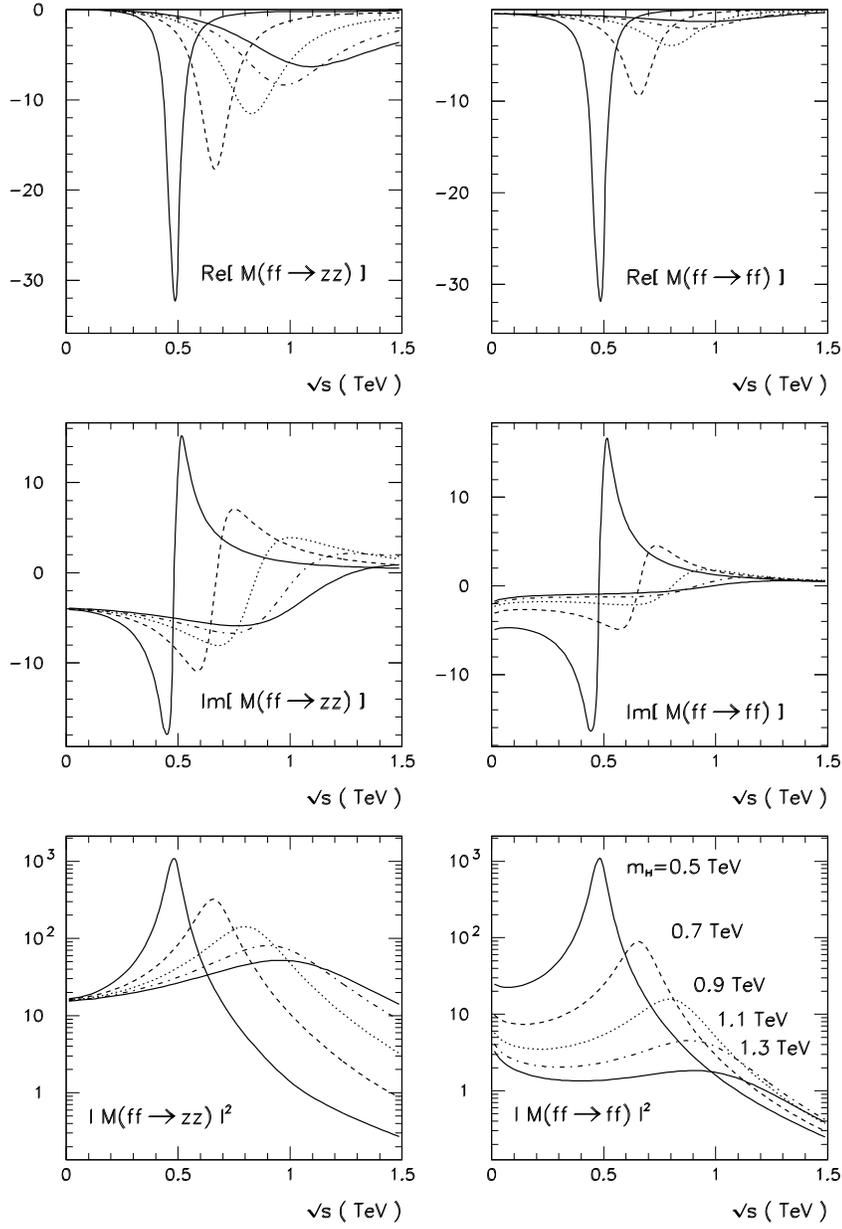}
\caption{\label{pawfiga}{\em Real, imaginary part and absolute square of the amplitude 
${\cal M}(f\bar f\rightarrow zz)$ (left) and ${\cal M}(f\bar f\rightarrow f'\bar f')$ (right).}}
\end{figure}

In figure \ref{pawfiga} we plot the amplitudes of the two processes
for several values of the coupling parameterized 
by $m_h$, defined in eq. \ref{epspre}.
The Yukawa coupling is set to one. Note how the Higgs signal evolves
from a sharp resonance to the damped regime, 
as the coupling is getting strong.   

\begin{figure}
\hspace{.5cm}
\epsfxsize = 12.5cm
\epsffile{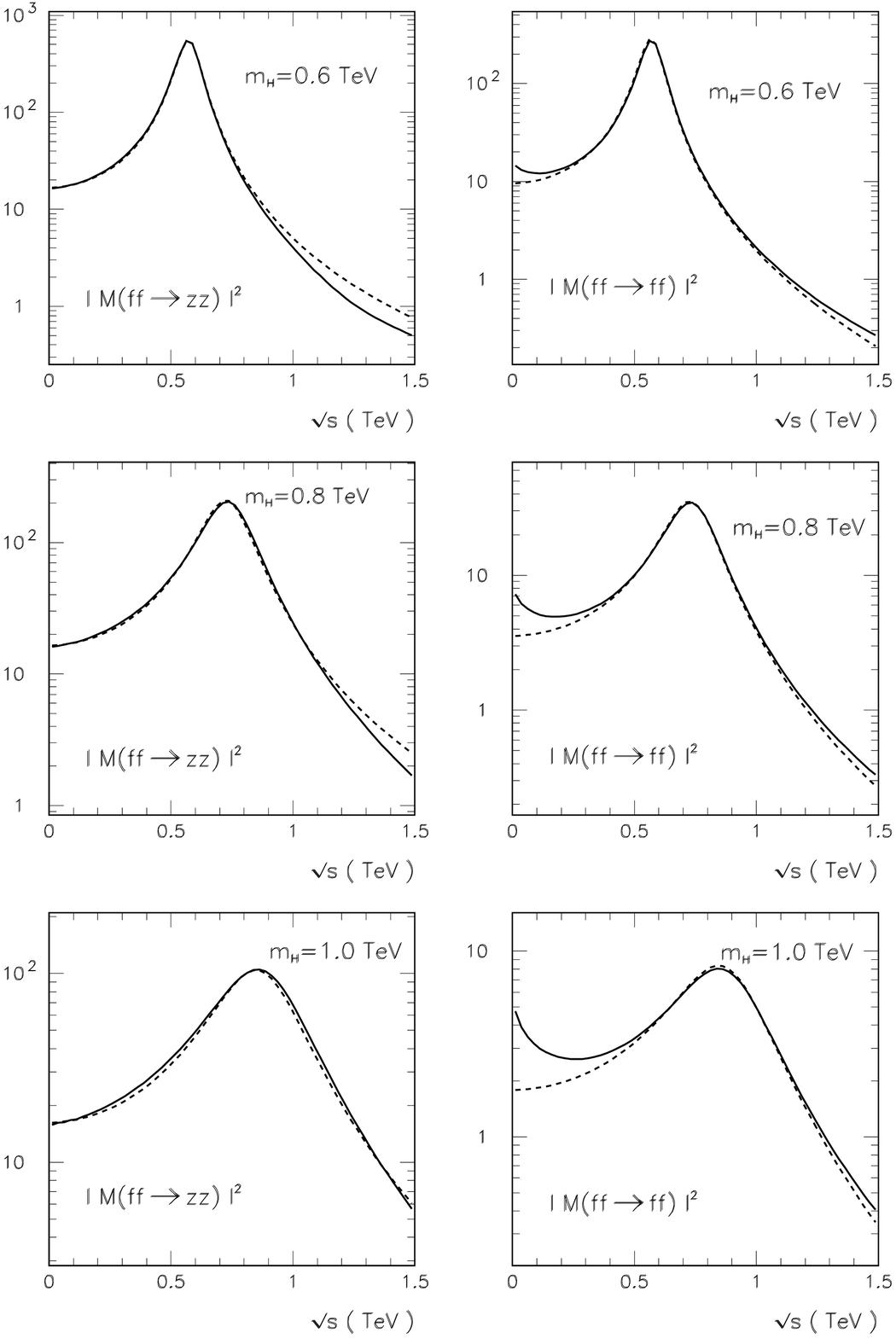}
\caption{\label{compare}{\em Comparison between the line shapes computed
in next-to-leading order $1/N$ (solid line) 
and two-loop perturbation theory (dashed line).} }
\end{figure}

We compare our results with the perturbative ones in figure \ref{compare}.

The peaks of the NLO $1/N$ and NNLO perturbation theory 
result are aligned to allow for a 
physically meaningful comparison between the two calculations. 
For on-shell Higgs masses below $1000$ GeV the curves
practically coincide. From this figure it is obvious that the results of 
both approaches agree with each other up to the highest 
accessible energies of the LHC. Away from the resonance, the perturbative 
curves which we give 
are not expected to be a good approximation. Close to the vector 
boson pair production threshold,
the equivalence theorem is not valid anymore, 
and the curves will be modified by 
contributions of order of the vector boson masses.

\begin{figure}
\hspace{.5cm}
\epsfxsize = 12.5cm
\epsffile{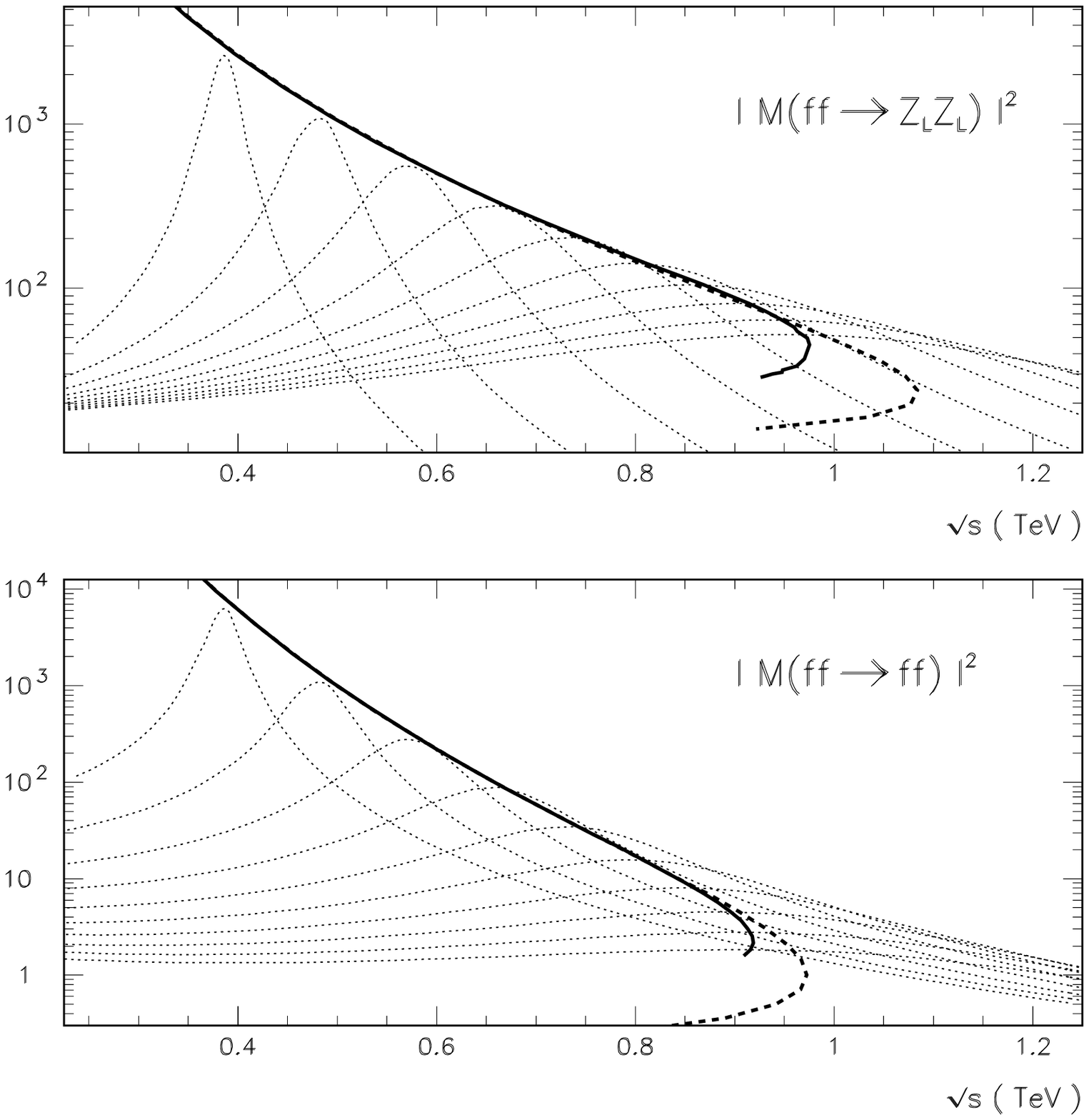}
\caption{\label{satur}{\em The saturation effect of 
the Higgs resonance for the processes
$f\bar f\rightarrow zz$ (a) and  
$f\bar f\rightarrow f'\bar f'$ (b). The solid and dashed 
curves are the position of the peak computed in the $1/N$ 
expansion and perturbation theory, respectively.} }
\end{figure}

As discussed in earlier publications \cite{1on:nlo},
one finds in both approaches a saturation of the peak position. 
This is confirmed now also for the process 
$f\bar f\rightarrow zz$. The saturation effect is shown in figure \ref{satur}.

It turns out that the NNLO perturbative result allows a maximal 
peak position of about 1100 GeV, whereas for the $1/N$ result 
one finds 975 GeV. 
In the case of $f\bar f\rightarrow f'\bar f'$ one finds 980 GeV for 
the perturbative and 930 GeV for the $1/N$ result. 
In this sense, the nonperturbative prediction is phenomenologically more
restrictive. 

\section{Conclusions}

We described in some detail the computation of Higgs production 
processes in the nonperturbative $1/N$ expansion for processes which
are relevant for future colliders.  
We want to stress that our calculation reproduces the perturbative results
in the region up to around $m_h=800$ GeV in a very precise manner,
which shows the consistency between perturbative methods
and  the $1/N$ expansion including
our tachyon subtraction scheme. Once the calculational tools
are developed, the calculation is not more difficult than a 
two-loop calculation in the Higgs sector.
Beyond $m_h \sim 900$ GeV, 
there are differences between the $1/N$ expansion
and perturbation theory concerning the position of the peak maxima,
but the implications for Higgs-discovery physics are not too large up to $\sim 1.1$ TeV.
This leads to the conclusion
that the Higgs sector is theoretically well understood 
for the entire kinematic 
region relevant for Higgs searches at LHC \cite{1on:lhc}.

A saturation of the peak position at strong coupling seems to
be a general feature of a heavy Higgs boson.

Note that the $1/N$ approach disentangles the uncertainties from the
renormalization scheme which is simply not present,
from the cutoff effects due to the tachyonic pole.
The effect of the tachyonic regularization
becomes more important as the coupling is increased, modeling 
nondecoupling effects from  a more fundamental theory 
containing the Higgs sector as an effective theory.


\section*{Appendix}
The following formulae \cite{ghinculov:3loop,denner:box} 
are used to get the integral representations 
of the $1/N$ graphs of eq. \ref{integralreps}.
With its help  the appearing two- and three-loop topologies 
can be cast into the form of one-loop integrals.

\subsection*{Three-point massless scalar integral:}                 

\begin{eqnarray}                    
C_0(k^2,p^2,q^2) =  \frac{1}{8 \pi^2}\frac{1}{k^2\sqrt{\Delta}}
 \Biggl[ Li_2\Bigl(-\frac{u_2}{v_1}\Bigr) + Li_2\Bigl(-\frac{v_2}{u_1}\Bigr) \qquad \qquad\qquad\nonumber \\
         + \frac{1}{4} \log^2\Bigl(\frac{u_2}{v_1}\Bigr)
         + \frac{1}{4} \log^2\Bigl(\frac{v_2}{u_1}\Bigr) 
         + \frac{1}{4} \log^2\Bigl(\frac{u_1}{v_1}\Bigr)  
         - \frac{1}{4} \log^2\Bigl(\frac{u_2}{v_2}\Bigr) + \frac{\pi^2}{6} \Biggr]
\end{eqnarray}
with
\begin{eqnarray} 
         u_{1,2} &=& (1+b-a\pm\sqrt{\Delta} )/2   \nonumber \\
         v_{1,2} &=& (1-b+a\pm\sqrt{\Delta} )/2   \nonumber \\
         \Delta  &=& 1 - 2 (a+b) + (a-b)^2        \nonumber \\
         a &=& p^2/k^2                            \nonumber \\
         b &=& q^2/k^2                            \nonumber 
\end{eqnarray}

\subsection*{Four-point massless scalar integral:} 

We use the notation $k_{ij}=-(\sum_{l=i}^{j-1} k_l)^2$
for the Mandelstam variables which are defined by the external momenta $k_{l=1,\dots,4}$:
\begin{eqnarray}  
D_0(k_1,k_2,k_3) = D_0(k_{12},k_{13},k_{14},k_{23},k_{24},k_{34}) = 
\hspace{5cm} \nonumber \\
\frac{1}{(4\pi)^2a (x_1-x_2)} \sum\limits_{k=1}^{2} (-1)^k \Biggl\{ -\frac{1}{2} \log^2(-x_k) 
\hspace{5cm}  \nonumber \\
 - Li_2\Bigl(1+\frac{k_{34}-i\delta}{k_{13}-i\delta} x_k \Bigr) 
   - \eta\Bigl(-x_k,\frac{k_{34}-i\delta}{k_{13}-i\delta} \Bigr)
       \log\Bigl(1+\frac{k_{34}-i\delta}{k_{13}-i\delta}x_k\Bigr) \nonumber \\
 - Li_2\Bigl(1+\frac{k_{24}-i\delta}{k_{12}-i\delta}x_k\Bigr) 
   - \eta\Bigl(-x_k,\frac{k_{24}-i\delta}{k_{12}-i\delta} \Bigr)
       \log\Bigl(1+\frac{k_{24}-i\delta}{k_{12}-i\delta} x_k\Bigr) \nonumber \\
  +  \log(-x_k) \Bigl[ \log(k_{12}-i\delta)+\log(k_{13}-i\delta)  \hspace{3cm}  \nonumber \\
                       -\log(k_{14}-i\delta)-\log(k_{23}-i\delta) \Bigr]\Biggr\}
                       \hspace{2cm}   
\end{eqnarray}                        
whereby,
\begin{eqnarray} 
x_{1,2} &=& \frac{-b\pm \sqrt{b^2-4 a c - 4 i a d \delta}}{2a} \nonumber \\
a &=& k_{24}k_{34} \nonumber \\
b &=& k_{13}k_{24} +k_{12}k_{34} - k_{14}k_{23} \nonumber \\
c &=& k_{12}k_{13}\nonumber \\
d &=& k_{23}\nonumber \\ 
\eta(u,v) &=& \log(uv) - \log(u) - \log(v)\nonumber 
\end{eqnarray} 
The cut of the logarithm lies along the negative real axis. An infinitesimal 
replacement $i\delta$ is included to stay on the correct Riemann sheet
if an argument vanishes. 

Both formulae allow for analytical continuation of an external momentum
which is important for our numerical purposes. This was checked
by recalculating several known two-loop diagrams  
with the same methods as used for the computation of  the 1/N graphs.



\newpage

\begin{thebibliography}{99}

\bibitem{kniehl:higgs1loop}  B. A. Kniehl, Phys. Rept. 240, (1993) 211.

\bibitem{marciano}           W.J. Marciano and S.S.D. Willenbrock, Phys. Rev. D37 (1988) 2509.

\bibitem{ghinculov:2loop}    A. Ghinculov and J.J. van der Bij, Nucl. Phys. {\bf B436} (1995) 30;  
                             A. Ghinculov, Nucl. Phys. B455 (1995)  21;
                                           Phys. Lett. B337 (1994) 137; (E) B346 (1995) 426.               
\bibitem{maher:2loop}  P. N. Maher, L. Durand and K. Riesselmann, Phys. Rev. D48 (1993) 1061;
                             (E) D52 (1995) 553; 
                             L. Durand, B.A. Kniehl and K. Riesselmann, Phys. Rev. D51 (1995) 5007;
                             Phys. Rev. Lett. 72 (1994) 2534; (E) Phys. Rev. Lett. 74 (1995) 1699; 
                             A. Frink, B.A. Kniehl, D. Kreimer, K. Riesselmann, 
                             Phys. Rev. D54 (1996) 4548.    

\bibitem{jikia:2loop}        V. Borodulin and G. Jikia; Phys. Lett. B391 (1997) 434.

\bibitem{degrassi:higgsbound} G. Degrassi, P. Gambino, M. Passera and A. Sirlin, 
                              Phys. Lett. B418 (1998) 209. 

\bibitem{lattice:higgsbound}  W. Langguth, I. Montvay, Z. Phys. C36 (1987) 725;
                              A. Hasenfratz, T. Neuhaus, Nucl. Phys. B297 (1988) 205.

\bibitem{coleman}     S. Coleman, R. Jackiw, H.D. Politzer, Phys. Rev. D10 (1974) 2491;

\bibitem{schnitzer}   H.J. Schnitzer, Phys. Rev. D10 (1974) 1800;
                      L. Dolan and R. Jackiw, Phys. ReV. D9 (1974) 3320.
                      
\bibitem{einhorn}     M.B. Einhorn, Nucl. Phys. B246 (1984) 75. 
                      R. Casalbuoni, D. Dominici and R. Gatto, Phys. Lett. B147 (1984) 419.

\bibitem{1on:nlo}     A. Ghinculov, T. Binoth, J.J. van der Bij, Phys. Rev. D57 (1998) 1487;
                      T. Binoth, A. Ghinculov, J.J. van der Bij, Phys. Lett. B417 (1998) 343. 


\bibitem{schnitzer:cutoff} J.P. Nunes, H.J. Schnitzer, Int.J.Mod.Phys. A10 (1995) 719. 

\bibitem{ghinculov:3loop} A. Ghinculov, Phys. Lett. B385 (1996) 279.

\bibitem{denner:box}  A. Denner, U. Nierste, R. Scharf, Nucl. Phys. B367 (1991) 637.   

\bibitem{1on:lhc}     A. Ghinculov, T. Binoth, J.J. van der Bij, Phys. Lett. B427 (1998) 343.


\end{thebibliography}
\end{document}